\documentclass[runningheads]{llncs}

\usepackage[T1]{fontenc}
\usepackage{graphicx}
\usepackage{booktabs}
\usepackage{amsmath}
\usepackage{amssymb}
\usepackage{hyperref}

\begin{document}

\title{TASR: Training-Free Adaptive Stopping for Iterative Retrieval}
\titlerunning{TASR: Training-Free Adaptive Stopping}

\author{Adrian Kieback\inst{1,2}\orcidID{0009-0002-8737-5974} \and
Uyiosa Philip Amadasun\inst{1,2}\orcidID{0009-0001-1420-6689} \and
Aman Chadha\inst{1}\orcidID{0000-0001-6621-9003} \and
Aaron Elkins\inst{1,2}\orcidID{0000-0001-5291-1023}}

\authorrunning{A. Kieback et al.}

\institute{James Silberrad Brown Center for AI \and
San Diego State University\\
\email{adrian.kieback@gmail.com} (corresponding author)}

\maketitle

\begin{abstract}
Iterative retrieval-augmented generation agents commonly overspend by continuing to retrieve after the model has converged on an answer, incurring calls that change neither the prediction nor the supporting evidence. Existing remedies learn a stopping policy from labeled trajectories, tying the decision to a trained component that requires retraining for each new model or task. We propose \textbf{TASR} (Training-Free Adaptive Stopping Rule), a one-line predicate that fires when the model repeats its previous-round normalized answer \emph{and} the isotonically calibrated logit margin exceeds $0.25$. No classifier or value head is learned; the threshold is fixed across all thirty-two (model, retriever, corpus) configurations we evaluate. On a $3$-model $\times$ $2$-dataset distractor grid, TASR retains $94.8\%$ of fixed-$k{=}5$'s macro F1 at $62.6\%$ of its calls and exceeds fixed-$k{=}3$ by $+3.42$ F1. The pattern holds on nine open-domain BM25 cells ($55.01$ F1 at $2.98$ calls vs.\ $54.33$ at $3.00$ for fixed-$k{=}3$) and, with calibration locked from the distractor split, on nine dense-retrieval cells across two retriever families, and on eight cells of a Nemotron-3-Ultra-550B production model, with zero significant regressions in any extension. The rule was selected from an exhaustive enumeration of $381$ candidate stopping rules on the canonical selection cell, where no alternative Pareto-dominates it. A signal-quality analysis shows that verbalized $1$--$5$ confidence collapses on RLHF-tuned models ($96.5\%$ of values equal $5$, entropy $0.182$ nats), while the logit margin achieves $40\times$ better class-conditional separation, grounding the design in a measurable model pathology. TASR is an auditable, training-free Pareto baseline for adaptive stopping in iterative retrieval. Code is publicly available \url{https://github.com/JSBAICenter/TASR}

\keywords{retrieval-augmented generation \and adaptive stopping \and multi-hop question answering \and training-free agents \and calibration}
\end{abstract}

\section{Introduction}
\label{sec:intro}

A Retrieval-Augmented Generation (RAG) Large Language Model (LLM) alternates between retrieving evidence and generating partial answers. Each retrieval requires a call to an external system (e.g., a vector store, BM25 index, or search API), followed by a call to the LLM. Both incur time, monetary, and energy costs, and inference is now a major and growing contributor to large-scale LLM energy consumption~\cite{luccioni2024power}. Reducing the average number of calls per query therefore lowers all three costs simultaneously.

A common failure mode of iterative retrieval is overspending: the agent continues retrieving after the model has already converged on an answer, incurring calls that change neither the prediction nor the supporting evidence. The standard remedy is to learn a stopping policy from labeled trajectories. Adaptive-RAG~\cite{jeong2024adaptive} trains a classifier to route questions to no-retrieval, single-step, or multi-step strategies. Stop-RAG~\cite{stoprag2024} learns a value head over retrieval trajectories, while Self-RAG~\cite{asai2024selfrag} trains the language model to emit retrieve and critique tokens. All three place the stopping decision in a learned component that is less transparent and typically requires retraining for new models or tasks.

This paper studies the stopping decision in isolation. We present \textbf{TASR} (Training-free Adaptive Stopping Rule), a one-line predicate that stops the loop when the model repeats its previous-round normalized answer \emph{and} the calibrated probability of being correct exceeds $0.25$. TASR uses no learned stopping controller: no classifier, value head, or policy trained on trajectories. Throughout, \emph{training-free} refers to the absence of such a learned decision policy, not the absence of any calibration; TASR's only fitted component is a one-dimensional per-round isotonic calibrator, estimated once from each model's $100$-question tune split and then frozen. The threshold is fixed across all thirty-two (model, retriever, corpus) configurations we evaluate. It serves as a drop-in alternative to fixed call budgets and learned controllers~\cite{trivedi2023interleaving,press2023measuring,jiang2023active,jeong2024adaptive}, transferring across the tested models and datasets without retraining.

Our contributions can be summarized as follows: (1)~a one-line predicate that beats fixed-$k{=}3$ by $+3.42$ macro F1 on a $3{\times}2$ distractor grid at matched cost; (2)~external validation on nine open-domain BM25 cells (HotpotQA-fullwiki, NQ-Open, TriviaQA-Open, all three models) and nine dense-retrieval cells across two retriever families, confirming that the pattern generalizes beyond the curated distractor pool and beyond lexical retrieval; (3)~full auditability using two scalars per round; (4)~a grounded analysis showing that Gemma's ${\sim}2\times$ overconfident raw logit margins are corrected by isotonic calibration, explaining why TASR's largest gains appear on the weakest model; (5)~a signal-quality diagnostic showing that verbalized confidence collapses on RLHF-tuned models while the logit margin preserves $40\times$ better class-conditional separation; and (6)~a scale stress test on Nemotron-3-Ultra-550B (${\sim}18$--$23\times$ the design scale) on which the rule and locked threshold transfer unchanged with zero significant regressions.

\section{Related Work}
\label{sec:related}

\paragraph{Iterative retrieval.} IRCoT~\cite{trivedi2023interleaving} interleaves chain-of-thought generation with retrieval, using each reasoning step to guide the next query; it improves multi-hop QA on HotpotQA, 2WikiMultiHopQA, MuSiQue, and IIRC, but issues a retrieval call per CoT sentence with no internal stopping criterion. Self-Ask~\cite{press2023measuring} decomposes a complex question into follow-up sub-questions, retrieving evidence for each. FLARE~\cite{jiang2023active} predicts the next sentence and triggers retrieval whenever it contains a low-confidence token, then regenerates; this is the closest prior signal to the calibrated logit margin we use, but FLARE applies it within a single generation pass and uses it to decide \emph{when to retrieve}, not when to stop. None of these works isolates the stopping decision as a standalone component.

\paragraph{Adaptive routing and learned stopping.} Adaptive-RAG~\cite{jeong2024adaptive} trains a smaller language model on automatically derived complexity labels to classify each incoming question as no-retrieval, single-step, or multi-step, and applies the strategy uniformly to the whole question. Self-RAG~\cite{asai2024selfrag} trains the generator language model itself to emit retrieve and critique special tokens, using a critic model supervised by GPT-4 to label training data. Stop-RAG~\cite{stoprag2024} trains a value head on full retrieval trajectories to decide when to halt. All three approaches tie the stopping decision to a trained component and require retraining per new model or task. TASR replaces the trained component with an equality check and a calibrated threshold.

\paragraph{Self-consistency and answer stability.} Self-consistency~\cite{wang2023selfconsistency} samples multiple independent reasoning chains at a fixed evidence set and selects the modal answer; agreement \emph{across samples} signals correctness. Our answer-stable component is the iteration-level analogue: agreement across consecutive rounds within a single trajectory.

\paragraph{Calibration and verbalized confidence.} Prior work shows that verbalized confidence is often poorly calibrated in RLHF-tuned language models~\cite{tian2023just}. This limits its usefulness as a decision signal and motivates the use of model-internal uncertainty measures instead. In our setting, we therefore use the logit gap at the answer-commitment token and calibrate it against correctness.

\paragraph{Evidence dilution.} Adding retrieved context can reduce performance through distracting evidence or context dilution~\cite{multidoc_rag_2025,multihop_dilution_2025}. For iterative retrieval systems, this implies that additional rounds are not uniformly beneficial.
\section{Method}
\label{sec:method}

\begin{figure}[!t]
\centering
\includegraphics[width=0.9\textwidth]{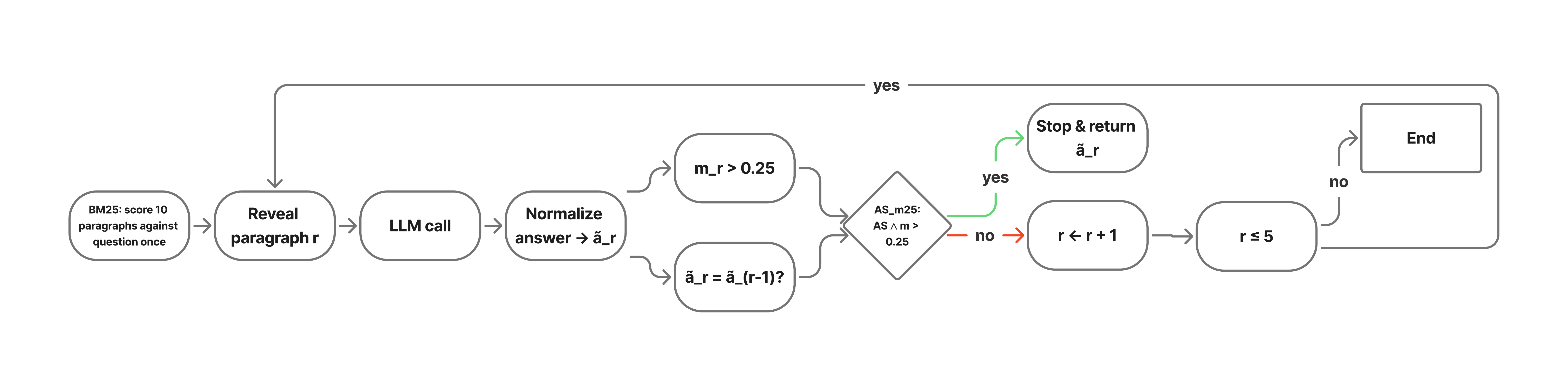}
\caption{The TASR loop. The retriever ranks candidate paragraphs against the question once; each round reveals the next-ranked paragraph and calls the LLM. The raw answer is normalized to $\tilde a_r$ and compared with the previous round's $\tilde a_{r-1}$. The first answer-token logit margin $m_r$ is isotonically calibrated on the tune split (offline) and compared against $0.25$. TASR fires when both checks pass; otherwise the loop continues until the $5$-round budget is exhausted. \S\ref{sec:rule} provides a worked example.}
\label{fig:loop}
\end{figure}

\subsection{Agent and Action Space}

The agent operates in a two-action setting. At each round $r \geq 1$ the policy chooses between \emph{stopping and answering} or \emph{adding the next ranked paragraph to the evidence set}. There is no query reformulation: BM25 (or the dense retriever) ranks candidate paragraphs against the question once, and each round reveals the next passage in that fixed ranked list. Round-$r$ state is exactly the top-$r$ retrieved paragraphs. One LLM call per round produces a structured response: a short answer preceded by the literal token ``Answer:'', a self-assessed $1$--$5$ confidence, and the per-token logprobs from which the margin signal is computed. The maximum number of rounds is $R{=}5$. If no stopping rule fires, the agent returns the round-$5$ answer.

We add one paragraph per round to maximize the resolution of the stopping signal. Coarser granularity is left to future work.

\subsection{Cached Signals}
\label{sec:signals}

Each round produces four signals that are cached to disk and used by any stopping rule. We group them by the sufficiency criterion they are intended to capture.

\paragraph{Answer stability (trajectory).} \texttt{answer\_stable} $\in \{0, 1\}$ is $1$ if the round-$r$ answer, normalized with the standard HotpotQA scoring procedure (lowercase, strip articles, strip punctuation, collapse whitespace), equals the round-$(r{-}1)$ normalized answer. It is defined for $r \geq 2$.

\begin{sloppypar}
\paragraph{Calibrated logit margin (model convergence).} \texttt{calibrated\_logit\_margin} $\in [0, 1]$ is the calibrated probability that the round-$r$ answer is exactly-match correct. The raw signal is \texttt{answer\_token\_margin}, the top-1 minus top-2 logprob at the first non-whitespace token after the literal string ``Answer:'' in the generated response, which is the point at which the model commits to its output. Calibration uses per-round isotonic regression to map  \texttt{answer\_token\_margin} to $P(\mathrm{EM}{=}1)$, fit on the $100$-question tune split for each cell. This signal preserves the model's uncertainty even when the verbalized $1$--$5$ confidence collapses to a single value, a known pathology on RLHF-tuned models~\cite{tian2023just}.
\end{sloppypar}

\paragraph{Lexical overlap (information novelty).} \texttt{overlap\_signal} $\in [0, 1]$ is the IDF-weighted token Jaccard between the round-$r$ paragraph and the union of paragraphs $1, \ldots, r{-}1$. High overlap means the new evidence is mostly redundant.

\paragraph{Verbalized confidence (model convergence).} \texttt{calibrated\_conf} $\in [0, 1]$ is the parsed $1$--$5$ self-reported confidence, isotonically calibrated against per-row exact match on the tune split.

All four signals are persisted to a single per-round parquet, so every stopping rule is evaluated offline against the cache with no fresh LLM calls. Only \texttt{answer\_stable} and \texttt{calibrated\_logit\_margin} enter the final rule. The other two signals are inputs to the search of \S\ref{sec:ablation}.

\begin{figure}[t]
  \centering
  \includegraphics[width=0.85\textwidth]{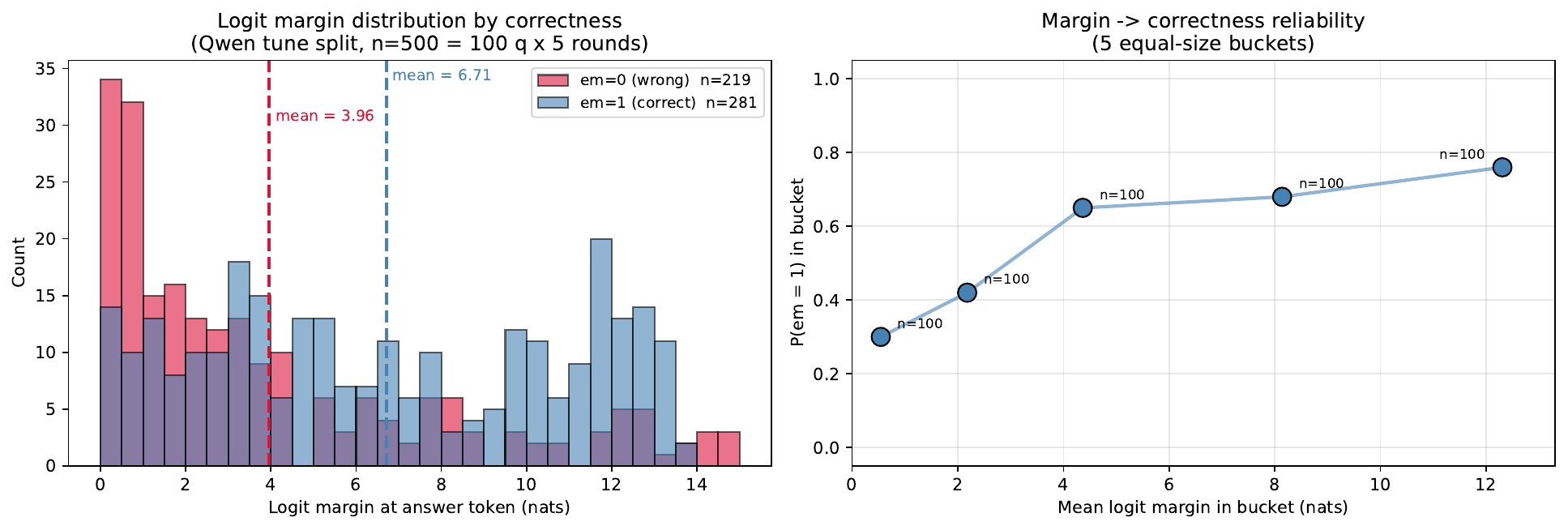}
  \caption{\textbf{Left:} Logit margin distribution by correctness on the Qwen HotpotQA-distractor tune split ($n{=}500$, $100$ questions $\times$ $5$ rounds). Correct answers (blue) have mean margin $6.71$ nats vs.\ $3.96$ for incorrect (red), a $2.75$-nat separation. \textbf{Right:} Reliability diagram: $P(\mathrm{EM}{=}1)$ in five equal-size margin buckets ($n{=}100$ each). The margin is monotonically informative.}
  \label{fig:margin}
  \end{figure}

\subsection{Per-Round Isotonic Calibration}
\label{sec:calibration}

The calibrated logit margin uses \emph{per-round} isotonic regression rather than a single global calibrator. This design choice is load-bearing. The empirical probability of correctness given the same confidence value varies substantially across rounds: on the Qwen tune split, $P(\mathrm{EM}{=}1 \mid \mathrm{conf}{=}5)$ rises from $0.442$ at round~$1$ to $0.643$ at round~$5$, a $20$-percentage-point swing reflecting the growing evidence set. A global calibrator averages away this round-dependent meaning.

\begin{sloppypar}
We fit one isotonic regression per round per cell using the pool-adjacent-violators (PAV) algorithm on the $100$-question tune split (each round contributes $100$ rows). The calibrator maps \texttt{answer\_token\_margin} $\to P(\mathrm{EM}{=}1)$ monotonically. The margin retains full spread at every round (round-4 margins span $0.19$ to $14.8$ nats on the Qwen tune split), so its calibrator never collapses to a trivial mapping. The verbalized confidence, by contrast, degenerates to a single value by round~$4$ ($100\%$ equal $5$), which is precisely why the rule relies on the margin rather than on confidence.

\end{sloppypar}

On a separate ablation across $18$ (model, dataset, prompt) cells, per-round isotonic achieves a worse Brier score than global isotonic ($0.195$ vs.\ $0.192$) but better rule-level F1 ($59.0$ vs.\ $57.9$). The reversal illustrates that standard calibration metrics (Brier) are not the right target for threshold-based stopping rules: a calibrator that minimizes average prediction error can blur the decision boundary at the threshold that matters for the rule (see \S\ref{sec:brier} for a fuller treatment of this phenomenon).

Platt scaling wins on $12$ of $18$ cells ($+0.41$ F1 avg) but is less robust when the margin distribution is narrow or multi-modal; we retain isotonic for robustness on small $n$.

\subsection{The TASR Predicate: AS\_m25}
\label{sec:rule}

TASR is the rule \textbf{AS\_m25}, defined as:
\begin{equation}
\mathrm{stop}_r \;=\; \mathbf{1}\!\left[\, \tilde{a}_r = \tilde{a}_{r-1} \;\wedge\; m_r > 0.25 \,\right],
\label{eq:asm25}
\end{equation}
where $\tilde{a}_r$ is the round-$r$ normalized answer and $m_r$ is the calibrated logit margin. In other words: stop at the first round at which the model repeats its previous-round answer and the calibrated probability of being correct exceeds $25\%$. By construction the rule cannot fire at $r{=}1$, so the minimum cost is two LLM calls per question.

The threshold $0.25$ is a single hand-set value, fixed within the pre-specified bracket $[0.20, 0.35]$ and locked across all thirty-two configurations with no per-cell tuning. As a post-hoc sensitivity check, sweeping the margin floor across this bracket on the distractor evaluation set keeps macro F1 within $[56.3, 58.4]$ at $3.0$ to $3.4$ calls, with every floor exceeding fixed-$k{=}3$ at matched cost, so the headline does not depend on the exact value. The only fitted component is a per-round isotonic calibrator, estimated once per model from its tune split and then frozen across all configurations.

The margin floor guards against wrong-and-stable answers: without it, the pure answer-stable rule fires on roughly $3\%$ of trajectories where the model repeats an incorrect answer before the gold paragraph enters, and the $0.25$ threshold converts roughly half of those premature stops into correct late-round answers.

By construction the rule cannot fire at $r{=}1$; a round-aware variant allowing an early stop saves $6.7\%$ of calls at an undetectable $-0.17$ F1 cost, which is not a meaningful Pareto extension.

\paragraph{Walk-through example.}
Figure~\ref{fig:loop} shows the loop structure. This paragraph traces one real trajectory through it. Consider HotpotQA dev question \texttt{5a77e70f}: \emph{``What is the title of
the 1979 film adaptation of William Shakespeare's play in which the English poet who wrote `Whale Nation' played a main character?''} (gold answer: \emph{The Tempest}). BM25 ranks the ten paragraphs once; the top two are ``Heathcote Williams'' (score $32.5$) and ``The Tempest (1979 film)'' (score $19.8$). At $r{=}1$ only the first paragraph is visible; the LLM answers \emph{``Titus Andronicus''} with calibrated margin $m_1{=}0.30$. A margin-only stopping rule ($m_r > 0.25$) would halt here on a wrong answer. AS\_m25 does not fire because answer-stable is undefined at $r{=}1$. At $r{=}2$ the gold paragraph ``The Tempest (1979 film)'' enters the evidence; the LLM revises to \emph{``The Tempest''} ($m_2{=}0.81$), but $\tilde{a}_2 \neq \tilde{a}_1$ so AS\_m25 still does not fire. At $r{=}3$ the model repeats \emph{``The Tempest''} ($m_3{=}0.80$): both predicates are satisfied and the loop stops, returning the correct answer in three calls rather than the fixed-$k{=}5$ budget.

\section{Experimental Setup}
\label{sec:setup}

\paragraph{Datasets.} \textbf{Distractor setting (primary).} \textbf{HotpotQA}~\cite{yang2018hotpotqa} distractor validation, sampled to $400$ questions with seed $42$ and split $100$ tune / $300$ eval. \textbf{2WikiMultiHopQA}~\cite{ho2020constructing} distractor validation, same sampling protocol. Both use the standard 10-paragraph pool ($2$ gold, $8$ distractor) per question.

\begin{sloppypar}
\textbf{Open-domain setting (external validity).} To test generalization beyond the curated distractor pool, we re-run the same pipeline on three open-domain BM25 corpora via prebuilt Pyserini Lucene indexes~\cite{lin2021pyserini}: \textbf{HotpotQA-fullwiki} (same $300$ eval questions, retrieved from the full Wikipedia index), \textbf{NQ-Open}~\cite{kwiatkowski2019natural} ($300$ questions subsampled with seed~$42$ from DPR's dev set), and \textbf{TriviaQA-Open}~\cite{joshi2017triviaqa} (same protocol). This yields $9$ additional cells: all three models on fullwiki, NQ, and TriviaQA.
\end{sloppypar}

\paragraph{Models.}
Three open-weight instruction-tuned LLMs served behind OpenAI-compatible endpoints at temperature $0$ with the top-$5$ logprobs returned per token.
Table~\ref{tab:models} summarizes the weight precision, KV cache dtype, and serving stack for each model.
The distractor grid contains six cells; the open-domain extension adds nine more (\S\ref{sec:external}). As a scale stress test (\S\ref{sec:scale}) we additionally evaluate \textbf{Nemotron-3-Ultra-550B}~\cite{nvidia_nemotron3_2026} ($55$B active
parameters, NVFP4 weights, fp8 KV cache, served via vLLM), a production model with roughly $18$--$23\times$ the total parameters of the three design models; it is not part of the design grid.

   \begin{table}[h]
   \caption{Model serving configurations. All endpoints expose an OpenAI-compatible
 API at temperature $0$ with top-$5$ logprobs.}
   \label{tab:models}
   \setlength{\tabcolsep}{0pt}
   \begin{tabular*}{\textwidth}{@{\extracolsep{\fill}} llccc @{}}
   \toprule
   Model & Ref. & Weight precision & KV cache & Server \\
   \midrule
   Qwen 3.6-27B & \cite{qwen_team_2024} & Q4\_K\_M (4-bit GGUF) & fp16 & llama.cpp
 \\
   Devstral-Small-2-24B-Instruct-2512 & \cite{mistral_devstral_2025} & fp8 & fp16 &
  vLLM \\
   Gemma-4-31B-IT & \cite{google_gemma4_2025} & fp8 & fp16 & vLLM \\
   Nemotron-3-Ultra-550B & \cite{nvidia_nemotron3_2026} & NVFP4 & fp8 & vLLM \\
   \bottomrule
   \end{tabular*}
   \end{table}

\paragraph{Baselines and caching.} Closed-book, fixed-$k$ for $k \in \{1, 2, 3, 5\}$, and a per-question oracle (earliest correct round). Fixed-$k{=}3$ is the primary budget baseline. All LLM responses are disk-cached, and every rule sweep is offline simulation on the cached parquet.

\paragraph{Cost model.} We report LLM call count rather than wall-clock latency, which depends on serving configuration outside our control; call count is reproducible and is exactly what TASR reduces. Call count is not identical to token cost, but the two move together here: each call is a single short-answer generation over the top-$r$ retrieved paragraphs, and the rounds TASR skips are the late, largest-context ones, so a saved round removes more tokens than an early round would. The incremental one-paragraph-per-round design is moreover compatible with prefix key-value caching, since round $r{+}1$ extends round $r$'s context by a single paragraph; we did not control caching on the third-party endpoints, so we report calls as the serving-independent cost metric.

\paragraph{Statistics.} Paired bootstrap with $1{,}000$ resamples and seed $42$, paired by question id within a cell~\cite{paired_bootstrap_2025}. Confidence intervals are the $2.5$ and $97.5$ percentiles of the resampled distribution.

\section{Results}
\label{sec:results}

\subsection{Headline}

Table~\ref{tab:headline} reports macro F1 and macro calls across the six distractor cells. TASR reaches $57.40$ F1 at $3.13$ calls, exceeding fixed-$k{=}3$ by $+3.42$ F1 at $+0.13$ calls and retaining $94.8\%$ of fixed-$k{=}5$'s F1 at $62.6\%$ of its calls. The per-question oracle reaches $65.26$ F1 at $1.61$ calls; it is below fixed-$k{=}3$ on calls because it can stop at round $1$ on questions the closed-book pass already answers correctly.

\begin{table}[h]
\caption{Macro results across the $3 \times 2$ grid ($n{=}6$ cells, $300$ eval questions per cell).}
\label{tab:headline}
\setlength{\tabcolsep}{5pt}
\begin{tabular}{lrrr}
\toprule
Method & macro F1 & macro calls & $\Delta$F1 vs.\ $k{=}3$ \\
\midrule
Fixed $k{=}1$         & $35.29$ & $1.00$ & $-18.69$ \\
Fixed $k{=}3$         & $53.98$ & $3.00$ & ref \\
\textbf{TASR (AS\_m25)} & $\mathbf{57.40}$ & $\mathbf{3.13}$ & $\mathbf{+3.42}$ \\
Fixed $k{=}5$         & $60.52$ & $5.00$ & $+6.54$ \\
\midrule
Oracle (UB)           & $65.26$ & $1.61$ & $+11.28$ \\
\bottomrule
\end{tabular}
\end{table}

\begin{figure}[t]
\centering
\includegraphics[width=0.49\textwidth]{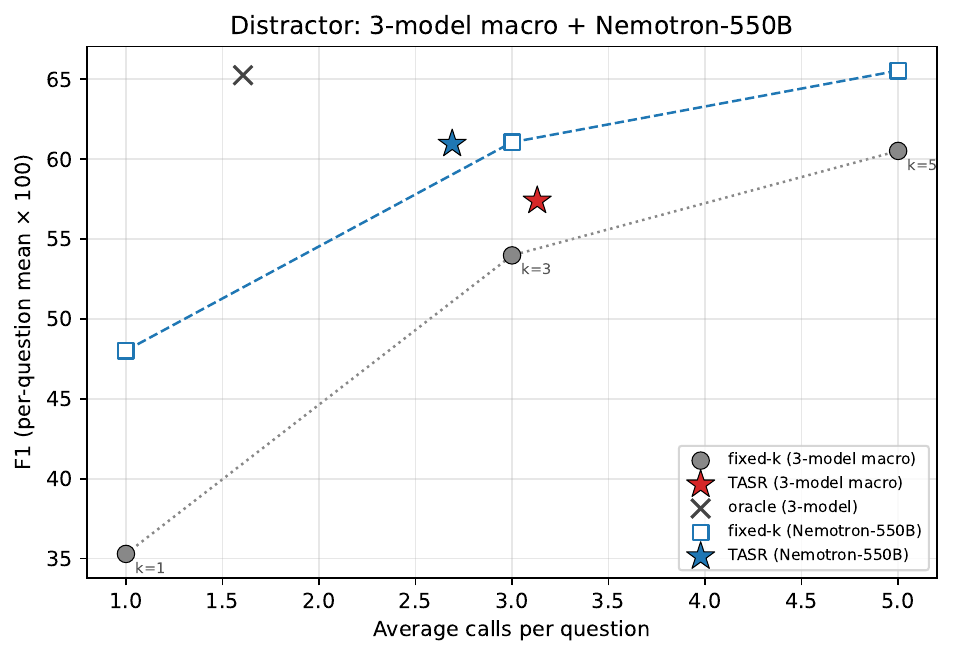}\hfill
\includegraphics[width=0.49\textwidth]{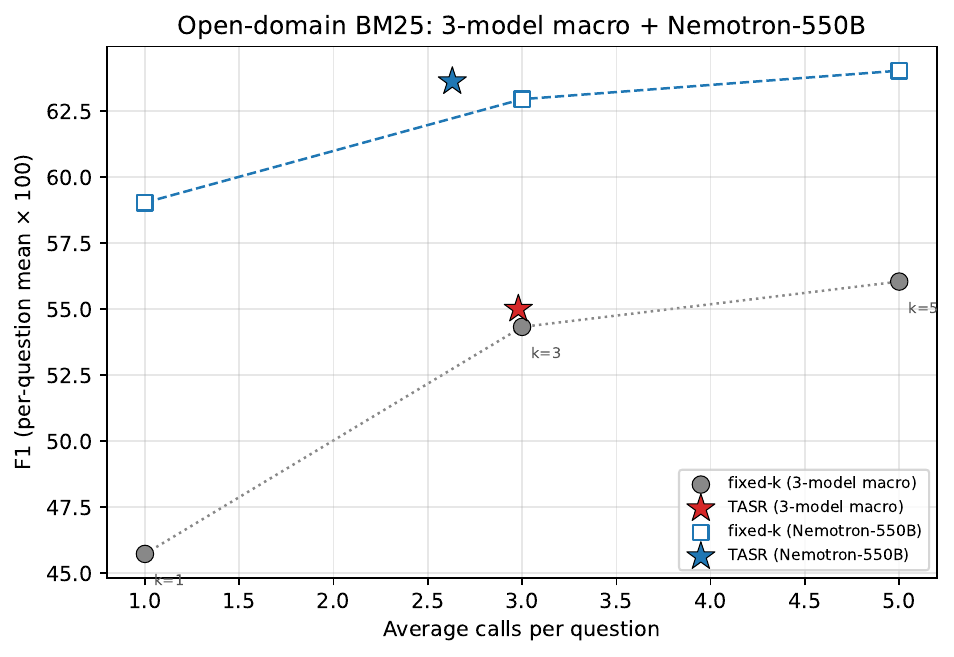}
\caption{F1 vs.\ calls. \textbf{Left:} distractor, $3$-model macro (red star) with the Nemotron-550B overlay. \textbf{Right:} open-domain BM25, $3$-model macro with the Nemotron-550B overlay. TASR sits above fixed-$k{=}3$ in both panels.}
\label{fig:pareto}
\end{figure}

Figure~\ref{fig:pareto} visualizes the F1-vs-calls Pareto frontier on both settings.

\subsection{Per-Cell Breakdown}

\begin{figure}[t]
\centering
\includegraphics[width=0.9\textwidth]{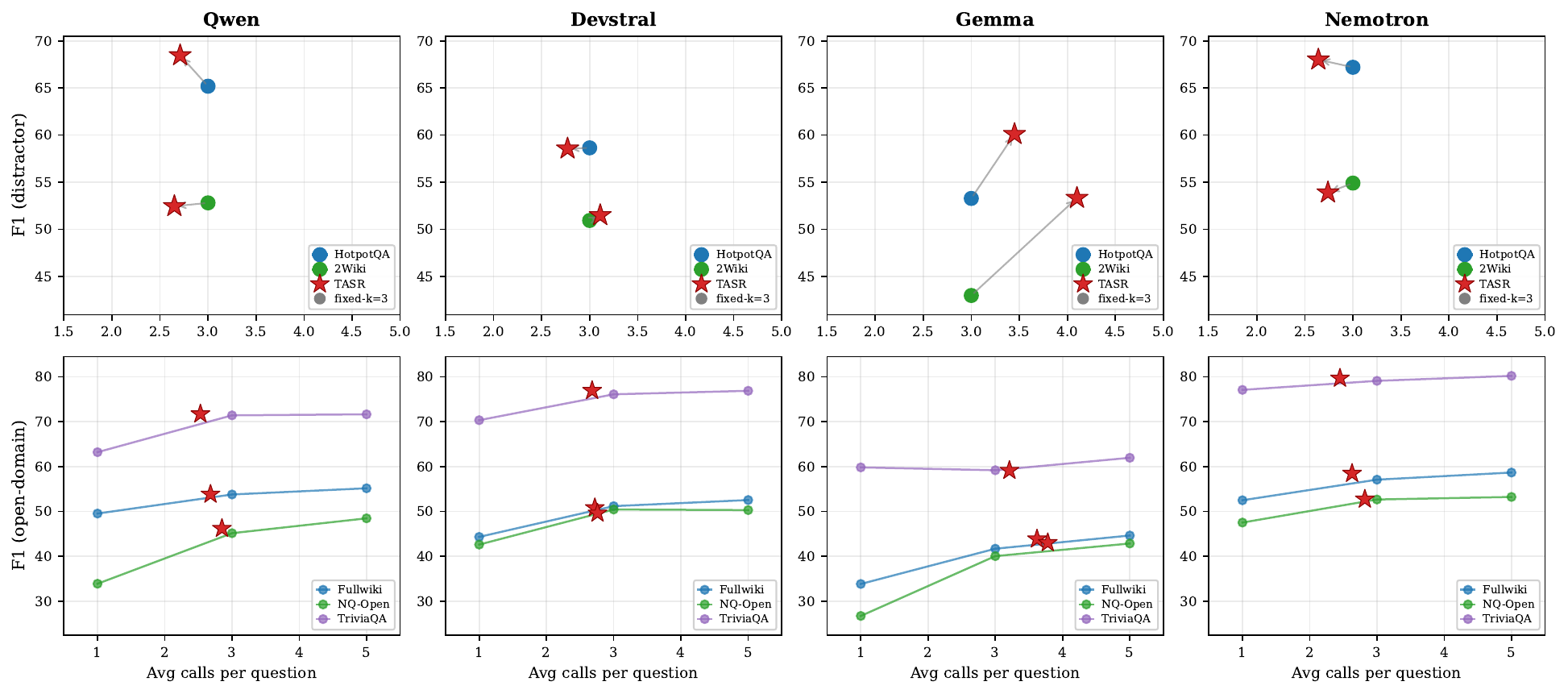}
\caption{Per-model Pareto breakdown. \textbf{Top:} Distractor cells (HotpotQA, 2Wiki). Blue/green circles = fixed-$k{=}3$; red stars = TASR. \textbf{Bottom:} Open-domain cells (lines = fixed-$k$ frontier; stars = TASR). Gemma shows the largest TASR gains; Devstral is the near-tie case. Nemotron-550B (fourth column, both rows) ties fixed-$k$ at fewer calls.}
\label{fig:permodel}
\end{figure}

Figure~\ref{fig:permodel} decomposes the macro Pareto plot by model and setting. The figure confirms three regimes: Qwen shows clean wins (arrows point up-left on HotpotQA), Devstral is a near-tie (arrows are flat), and Gemma shows the largest gains (large upward arrows on both distractor datasets; on open-domain TASR nearly matches fixed-$k{=}5$ at fewer calls). The per-model view supports the Gemma overconfidence analysis of \S\ref{sec:gemma}, showing that the model with the worst single-pass accuracy benefits most from extra rounds guarded by a calibrated margin. The Nemotron-550B panel shows the same pattern at scale.

Table~\ref{tab:percell} reports the TASR rule against fixed-$k{=}5$ and against the budget-matched fixed-$k{=}3$ baseline in each cell. Three cells show a statistically significant lift over fixed-$k{=}3$ (qwen\_hp, gemma\_hp, gemma\_2w), while the other three (qwen\_2w, devstral\_hp, devstral\_2w) are statistical ties. No cell shows a significant regression. The macro lift is therefore an aggregate of strict wins and ties, rather than a mixture of wins and regressions.

\begin{table}[h]
\caption{Per-cell TASR results. $\Delta$F1 vs.\ fixed-$k{=}3$; bold CIs exclude zero.}
\label{tab:percell}
\setlength{\tabcolsep}{3pt}
\scriptsize
\begin{tabular}{lrrrl}
\toprule
Cell & TASR F1 & calls & $\Delta$F1 & 95\% CI \\
\midrule
qwen\_hp      & $68.46$ & $2.71$ & $\mathbf{+3.27}$ & \textbf{[+1.48, +5.30]} \\
qwen\_2w      & $52.45$ & $2.65$ & $-0.36$ & [$-3.16, +2.63$] \\
devstral\_hp  & $58.56$ & $2.77$ & $-0.09$ & [$-1.79, +1.93$] \\
devstral\_2w  & $51.49$ & $3.11$ & $+0.55$ & [$-2.96, +4.37$] \\
gemma\_hp     & $60.10$ & $3.45$ & $\mathbf{+6.82}$ & \textbf{[+4.21, +10.00]} \\
gemma\_2w     & $53.33$ & $4.10$ & $\mathbf{+10.35}$ & \textbf{[+6.83, +14.03]} \\
\midrule
\textbf{macro} & $\mathbf{57.40}$ & $\mathbf{3.13}$ & $\mathbf{+3.42}$ & n/a \\
\bottomrule
\end{tabular}
\end{table}

\subsection{The Exhaustive Search Behind AS\_m25}
\label{sec:ablation}

AS\_m25 is the product of an exhaustive combinatorial search. On the canonical HotpotQA-distractor setting (Qwen), we enumerated every non-empty OR-combination of eight per-round signals (three of the four \S\ref{sec:signals} signals plus five computed solely for this search): \texttt{answer\_stable}, calibrated verbalized confidence, IDF-weighted lexical overlap, the BM25 top-score $z$-score, the rank-$1$/$2$ gap, an answer-in-evidence indicator, a cross-encoder reranker score (BGE-reranker-v2-m3~\cite{chen2024bge}), and an LLM-judge entailment score. This yielded $2^8\!-\!1\,{=}\,255$ candidate rules; a second pass on a confidence-first prompt with six of these signals added $2^6\!-\!1\,{=}\,63$ more ($318$ OR-subsets). Of these, only two produced a $95\%$ CI excluding zero against fixed-$k{=}3$, and both were single-signal rules: answer-stability under the canonical prompt, and calibrated verbalized confidence under the confidence-first prompt. We then added the calibrated logit margin (\S\ref{sec:rule}) and re-ran the enumeration over six signals ($63$ more), from which AS\_m25 emerged as the lone significant winner, for $381$ enumerated rules in total. The same locked threshold is then applied unchanged to all thirty-two configurations with no per-cell tuning.

Because we screen $381$ candidate rules, single-cell significance is suggestive rather than confirmatory; the frozen evaluation on the remaining thirty-one configurations, on which no rule is re-selected, is the actual out-of-sample guard against multiple-comparison bias.

Across all threshold sweeps, reranker variants, and several hundred further evaluations on this canonical cell, no rule Pareto-dominates AS\_m25. Two structural patterns explain why. First, OR-combining signals accelerates stopping in the wrong direction: each individually-tuned signal fires on roughly $50\%$ of rounds, so an OR of two fires on ${\sim}75\%$, triggering stops before evidence has accumulated. Second, AND-combining collapses to the strongest single constraint, which is AS\_m25 itself. The rule's leverage lies in its \emph{form} (a two-term AND-gate) rather than in the size of the signal set.

Five-fold cross-validation within each cell's $100$-question tune split confirms that per-cell tuning of the margin threshold does not pay off at this sample size: at a matched call budget (so that tuning cannot trade extra calls for F1), the locked threshold $0.25$ is $+0.94$ macro F1 better than per-cell-tuned margins, winning or tying on five of six cells. Locking therefore improves accuracy, not just transparency.

\subsection{Negative Results On Alternative Signals}
\label{sec:negative}

\paragraph{Cross-encoder reranker.} Adding a cross-encoder reranker (BGE-reranker-v2-m3~\cite{chen2024bge}) as a fifth OR-vote drops per-cell F1 by $4$--$14$ points (all CIs exclude zero). The reranker fires at round~$1$ on nearly every question because it scores evidence topicality rather than support for the model's current answer, short-circuiting the rule on still-wrong rounds. A per-question $z$-score variant is a flat null.

\paragraph{Continuous text stability.} Relaxing the binary check to a continuous token-F1 or SequenceMatcher threshold gains at most $+0.10$ F1 (CI $[0.00, +0.38]$) across all thresholds and cells. At temperature $0$ with HotpotQA normalization, the model produces highly stable text and there are no near-miss answers to recover.

\paragraph{BM25 retrieval confidence.} A robust $z$-score of the top BM25 score (median/IQR normalization) achieves $+0.568$ nats of class-conditional separation between correct and incorrect rows, a substantial improvement over the original global $z$-score ($-0.089$ nats, which is reverse-correlated). As a stopping rule, however, adding it to AS\_m25 as an OR-clause \emph{hurts} F1 by $3.5$--$17.9$ points across all thresholds tested. The retrieval-confidence signal fires too eagerly: it triggers stops on questions where BM25 is unambiguous but the LLM still gets the answer wrong. LLM-side signals dominate retriever-side signals for stopping.

\paragraph{Distribution margin aggregations.} On a 50-question pilot, replacing the first-answer-token margin with distribution-level aggregations (min, 10th-percentile, mean, median across all answer tokens) improves feature-level Pearson correlation with EM by up to $+57\%$ ($r = 0.60$ for \texttt{margin\_p10} vs.\ $0.38$ for \texttt{margin\_first}). At the rule level, however, substituting any aggregation into AS\_m25 changes macro F1 by at most $\pm 0.22$, within noise. The AND-gate with answer-stability is the binding constraint; margin improvements are wasted when the gate is closed.

\subsection{Signal Robustness: Tune vs.\ Evaluation}
\label{sec:dgap}

Table~\ref{tab:dgap} reports per-signal class-conditional Cohen's $d$ (correct vs.\ incorrect rows on the EM metric) on both the tuning split ($100$ questions) and the held-out evaluation split ($300$ questions), computed across all three models on the HotpotQA distractor cell, with Qwen shown as the representative case. Among the signals tested, \texttt{answer\_stable} has the highest eval $d$ ($0.481$) and a positive tune-to-eval delta ($+0.086$): its class-conditional separation \emph{strengthens} on the held-out split. The two-pass entailment judge shows the opposite pattern, losing more than half its separation from tune ($0.434$) to evaluation ($0.183$).

\begin{table}[h]
\caption{Class-conditional Cohen's $d$ (correct vs.\ incorrect on EM) by split. Positive $\Delta$ means the signal strengthens on evaluation. Qwen shown as representative; Devstral and Gemma show the same ordering.}
\label{tab:dgap}
\setlength{\tabcolsep}{4pt}
\scriptsize
\begin{tabular}{lrrr}
\toprule
Signal & tune $d$ & eval $d$ & $\Delta$ \\
\midrule
\textbf{answer\_stable} (headline) & 0.395 & \textbf{0.481} & $\mathbf{+0.086}$ \\
ce\_top\_score (reranker) & 0.106 & 0.268 & $+0.162$ \\
answer\_in\_evidence & 0.458 & 0.404 & $-0.054$ \\
calibrated\_conf (1--5) & 0.391 & 0.271 & $-0.120$ \\
overlap\_signal & 0.234 & 0.288 & $+0.054$ \\
top\_score\_z\_intra & 0.024 & 0.130 & $+0.106$ \\
gap (rank1$-$rank2) & 0.092 & 0.024 & $-0.068$ \\
\textbf{judge\_entailment} & \textbf{0.434} & \textbf{0.183} & $\mathbf{-0.251}$ \\
\bottomrule
\end{tabular}
\end{table}

Tune-set separation alone is a poor selector for stopping signals: the entailment judge looks strong on tune yet collapses on evaluation and doubles per-round LLM calls, whereas \texttt{answer\_stable} strengthens. This stability reflects the structure of trajectory convergence rather than a dataset-specific artifact. This generalizes across models: the AS\_m25 rule score (the full rule's calibrated quantity, distinct from the \texttt{answer\_stable} signal in Table~\ref{tab:dgap}) achieves eval $d$ of $0.55$--$1.00$ across all three models, with tune-to-eval drift $\leq 0.09$.

\subsection{External Validity: Open-Domain BM25 Corpora}
\label{sec:external}

To test whether TASR generalizes beyond the curated distractor pool, we re-run the identical pipeline on nine open-domain cells using Pyserini Lucene indexes, with calibration locked from each model's HotpotQA-distractor tune split. Table~\ref{tab:opendomain} reports the results.

\begin{table}[h]
\caption{Open-domain BM25 results ($300$ eval Qs/cell). Calibration locked from distractor tune split. $\Delta$F1 \& CIs: paired bootstrap ($1{,}000\times$, seed $42$) vs.\ $k{=}3$.}
\label{tab:opendomain}
\setlength{\tabcolsep}{3pt}
\scriptsize
\begin{tabular}{lrrrrl}
\toprule
Cell & $k{=}3$ F1 & $k{=}5$ F1 & TASR F1 / calls & $\Delta$F1 & 95\% CI \\
\midrule
qwen\_fullwiki     & $53.78$ & $55.16$ & $53.83$ / $2.68$ & $+0.05$ & [$-2.28, +1.95$] \\
qwen\_nq           & $45.15$ & $48.46$ & $46.20$ / $2.85$ & $+1.05$ & [$-1.25, +3.34$] \\
qwen\_trivia       & $71.42$ & $71.65$ & $71.72$ / $2.53$ & $+0.30$ & [$-1.63, +2.07$] \\
devstral\_fullwiki & $51.18$ & $52.54$ & $50.81$ / $2.72$ & $-0.38$ & [$-2.31, +1.34$] \\
devstral\_nq       & $50.43$ & $50.30$ & $49.61$ / $2.76$ & $-0.81$ & [$-2.39, +0.77$] \\
devstral\_trivia   & $76.11$ & $76.89$ & $76.94$ / $2.68$ & $+0.84$ & [$-0.63, +2.44$] \\
gemma\_fullwiki    & $41.69$ & $44.63$ & $43.84$ / $3.62$ & $+2.15$ & [$-0.27, +4.50$] \\
gemma\_nq          & $40.04$ & $42.85$ & $\mathbf{43.02}$ / $3.78$ & $\mathbf{+2.98}$ & \textbf{[+0.92, +5.29]} \\
gemma\_trivia      & $59.18$ & $61.96$ & $59.13$ / $3.21$ & $-0.06$ & [$-2.11, +2.12$] \\
\midrule
\textbf{macro} & $54.33$ & $56.05$ & $\mathbf{55.01}$ / $\mathbf{2.98}$ & $\mathbf{+0.68}$ & n/a \\
\bottomrule
\end{tabular}
\end{table}

The distractor-grid pattern is preserved: TASR matches fixed-$k{=}3$ F1 at matched cost and approaches fixed-$k{=}5$ at $40\%$ fewer calls. Six of nine cells show TASR $\geq$ fixed-$k{=}3$; the three cells with $\Delta$F1 $<0$ are devstral\_fullwiki ($-0.38$), devstral\_nq ($-0.81$), and gemma\_trivia ($-0.06$). The open-domain results are notable because the calibration is \emph{transferred}: per-model isotonic regression is fit once on each model's HotpotQA-distractor tune split and applied unchanged to entirely different corpora (fullwiki, NQ-Open, TriviaQA-Open), yet the Pareto pattern is preserved, suggesting the margin-to-correctness mapping is a property of the model rather than the corpus. The lift also concentrates on the weakest model (Gemma), so beyond cutting calls TASR delivers its largest accuracy gains where single-pass accuracy is lowest, reinforcing the efficiency motivation of \S\ref{sec:intro}.

\subsection{Retriever Robustness: Dense Retrieval, Two Families}
\label{sec:dense}

\begin{sloppypar}
To rule out that TASR's signal is BM25-specific, we re-run all three open-domain corpora with dense retrievers and locked calibration: \textbf{Contriever-MSMARCO}~\cite{izacard2022unsupervised} on HotpotQA-fullwiki and NQ-Open, and \textbf{DKRR-TQA}~\cite{izacard2021distilling} on TriviaQA-Open (DKRR is the standard dense baseline for that corpus, as Contriever has no prebuilt index for wikipedia-dpr-100w). The two families differ in architecture, training objective, and training corpus.
\end{sloppypar}

\begin{table}[h]
\caption{Dense retrieval, $9$ cells across two retriever families. Same $300$ eval Qs/cell; calibration locked from BM25-distractor tune split. $\Delta$F1 \& CIs vs.\ fixed-$k{=}3$.}
\label{tab:dense}
\setlength{\tabcolsep}{3pt}
\scriptsize
\begin{tabular}{lrrrrl}
\toprule
Cell & $k{=}3$ F1 & $k{=}5$ F1 & TASR F1 / calls & $\Delta$F1 & 95\% CI \\
\midrule
\multicolumn{6}{l}{\emph{Contriever-MSMARCO} \, $\cdot$\, HotpotQA-fullwiki} \\
qwen     & $51.10$ & $52.82$ & $51.10$ / $2.87$ & $+0.00$ & [$-1.84, +2.01$] \\
devstral & $43.60$ & $44.97$ & $43.92$ / $3.02$ & $+0.32$ & [$-1.30, +2.13$] \\
gemma    & $39.23$ & $42.33$ & $41.50$ / $3.84$ & $+2.27$ & [$-0.22, +4.81$] \\
\multicolumn{6}{l}{\emph{Contriever-MSMARCO} \, $\cdot$\, NQ-Open} \\
qwen     & $54.75$ & $57.11$ & $54.17$ / $2.71$ & $-0.58$ & [$-2.05, +0.70$] \\
devstral & $50.28$ & $50.35$ & $50.91$ / $3.04$ & $+0.63$ & [$-0.76, +2.10$] \\
gemma    & $43.31$ & $45.67$ & $\mathbf{46.57}$ / $3.60$ & $\mathbf{+3.26}$ & \textbf{[+1.34, +5.32]} \\
\multicolumn{6}{l}{\emph{DKRR-TQA} \, $\cdot$\, TriviaQA-Open} \\
qwen     & $76.97$ & $77.46$ & $75.89$ / $2.53$ & $-1.08$ & [$-2.79, +0.45$] \\
devstral & $74.40$ & $74.93$ & $73.61$ / $2.64$ & $-0.79$ & [$-2.42, +0.70$] \\
gemma    & $66.08$ & $67.27$ & $\mathbf{68.67}$ / $3.34$ & $\mathbf{+2.59}$ & \textbf{[+0.17, +5.07]} \\
\bottomrule
\end{tabular}
\end{table}

The Pareto pattern survives both swaps. Across all nine cells, no cell regresses significantly; two are significant wins (both Gemma: NQ-Open $+3.26$, TriviaQA-Open $+2.59$), and seven are ties. Both wins are on the model that gains most on BM25 (\S\ref{sec:gemma}): the retriever changes while the model-side failure mode the rule corrects does not. Calibration transfer is the key test: the isotonic was fit on BM25 logit margins from one corpus and applied unchanged to dense evidence from two retriever families on three corpora, TASR calls stay in $2.5$--$3.8$ across all nine cells, matching the BM25 range ($2.5$--$3.8$), so the calibrator did not silently fail.

\subsection{Scale Robustness: A 550B Production Model}
\label{sec:scale}

To test whether the rule survives a step change in model \emph{capability} rather than only in corpus or retriever, we re-run eight cells (two in-distribution distractor, three open-domain BM25, three open-domain dense) on Nemotron-3-Ultra-550B ($55$B active, NVFP4, served via vLLM)~\cite{nvidia_nemotron3_2026}, a production model with roughly $18$--$23\times$ the total parameters of the three design models. The rule, the $0.25$ threshold, and the frozen per-round isotonic procedure are unchanged; the only model-specific component is Nemotron's own calibrator, fit once on its $100$-question distractor tune split and never refit on eval data. Table~\ref{tab:nemotron} reports the results.

\begin{table}[h]
\caption{Scale stress test: Nemotron-3-Ultra-550B ($55$B active, NVFP4, vLLM). Calibration locked from its own $100$-question distractor tune split. $\Delta$F1 \& CIs: paired bootstrap ($1{,}000\times$, seed $42$) vs.\ fixed-$k{=}3$.}
\label{tab:nemotron}
\setlength{\tabcolsep}{3pt}
\scriptsize
\begin{tabular}{llrrrrl}
\toprule
Retriever & Corpus & $k{=}3$ & $k{=}5$ & TASR F1 / calls & $\Delta$F1 & 95\% CI \\
\midrule
\multicolumn{7}{l}{\emph{Distractor} (in-distribution)} \\
BM25       & HotpotQA & $67.20$ & $70.28$ & $68.01$ / $2.64$ & $+0.81$ & [$-1.68, +3.18$] \\
BM25       & 2Wiki    & $54.91$ & $60.80$ & $53.90$ / $2.74$ & $-1.01$ & [$-3.53, +1.47$] \\
\multicolumn{7}{l}{\emph{BM25} (lexical)} \\
BM25       & fullwiki & $57.08$ & $58.67$ & $58.47$ / $2.63$ & $+1.39$ & [$-0.70, +3.28$] \\
BM25       & NQ-Open  & $52.66$ & $53.23$ & $52.73$ / $2.82$ & $+0.07$ & [$-1.86, +1.77$] \\
BM25       & TriviaQA & $79.11$ & $80.22$ & $79.69$ / $2.45$ & $+0.58$ & [$-1.18, +2.32$] \\
\multicolumn{7}{l}{\emph{Dense} (neural)} \\
Contriever & fullwiki & $47.28$ & $50.40$ & $48.93$ / $2.85$ & $+1.65$ & [$-0.73, +4.03$] \\
Contriever & NQ-Open  & $40.87$ & $41.43$ & $41.11$ / $3.33$ & $+0.24$ & [$-1.60, +1.88$] \\
DKRR       & TriviaQA & $72.85$ & $74.73$ & $\mathbf{75.33}$ / $2.52$ & $\mathbf{+2.49}$ & \textbf{[+0.57, +4.51]} \\
\bottomrule
\end{tabular}
\end{table}

In-distribution, AS\_m25 ties fixed-$k{=}3$ on both distractor cells (HotpotQA $+0.81$, 2Wiki $-1.01$; both CIs span zero) at $2.64$ and $2.74$ calls. Under BM25, TASR significantly beats fixed-$k{=}1$ on all three corpora ($+2.6$ to $+6.0$ F1) and ties both fixed-$k{=}3$ and fixed-$k{=}5$ everywhere at $2.45$--$2.82$ calls, a $44$--$51\%$ reduction versus $k{=}5$ at no measurable accuracy cost. Under dense retrieval the pattern holds with zero significant regressions, and on TriviaQA (DKRR) the rule is a strict Pareto win: its $75.33$ F1 exceeds even fixed-$k{=}5$ ($74.73$) at $2.52$ calls. The stopping behavior is therefore invariant across more than an order of magnitude of model scale.

\subsection{Cross-Model Analysis: Why Gemma Benefits Most}
\label{sec:gemma}

Gemma is the weakest model on every cell yet receives the largest TASR lift. Table~\ref{tab:gemma_margins} reports four converging lines of evidence for a grounded hypothesis: Gemma combines a weak single-pass baseline with overconfident raw logit margins, and the per-round isotonic calibration is what corrects this.

\begin{table}[h]
\caption{Cross-model evidence for Gemma's overconfidence pattern. Raw and calibrated margins are means across each model's open-domain cells. $\Delta$F1 is TASR minus fixed-$k{=}3$, macro across respective cells.}
\label{tab:gemma_margins}
\setlength{\tabcolsep}{4pt}
\small
\begin{tabular}{lrrr}
\toprule
 & Qwen & Devstral & Gemma \\
\midrule
fixed-$k{=}1$ F1 (open-domain macro) & $48.87$ & $52.78$ & $35.54$ \\
raw margin (nats, open-domain mean) & $4.66$ & $3.71$ & $9.50$ \\
calibrated margin (open-domain mean) & $0.52$ & $0.49$ & $0.42$ \\
$\Delta$F1 vs.\ $k{=}3$ (distractor macro) & $+1.46$ & $+0.23$ & $+8.59$ \\
$\Delta$F1 vs.\ $k{=}3$ (open-domain macro) & $+0.47$ & $-0.12$ & $+1.69$ \\
\bottomrule
\end{tabular}
\end{table}

Gemma's single-pass baseline is $13$--$17$ F1 points below Qwen and Devstral, yet its raw logit margins are ${\sim}2\times$ larger, indicating overconfidence. Isotonic calibration reverses the ordering (Gemma lowest at $0.42$), mapping each model's margins onto a common $P(\mathrm{EM})$ scale so one locked threshold works across models. Because TASR spends more rounds on Gemma's less-stable trajectories (avg $3.21$--$3.78$ calls vs.\ $2.53$--$2.85$ for Qwen), the extra rounds buy real F1: $+8.59$ on distractor, $+1.69$ on open-domain, against $+1.46$/$+0.47$ for Qwen.

\subsection{Cross-Prompt Robustness}
\label{sec:crossprompt}

We re-ran the rule sweep on a \emph{confidence-first} prompt variant in which the model emits its $1$--$5$ confidence before committing to an answer. The reorder spreads the confidence distribution (entropy increases from $0.182$ to $0.839$ nats; the canonical prompt emits ``confidence $= 5$'' on $96.5\%$ of rounds) at the cost of $5$--$7$ F1 in raw accuracy (round-5 F1 drops from $69.4$ to $63.8$): the model is more willing to abstain when confidence precedes commitment. Under the confidence-first prompt, the headline shifts: the rule \texttt{calibrated\_conf} $> 0.552$ wins at $+5.84$ F1 $[+2.96, +8.86]$ vs.\ fixed-$k{=}3$, while \texttt{answer\_stable} drops to a tie ($-0.26$ F1 $[-2.60, +1.91]$).

Two consequences. (i)~The \emph{form} of the headline rule generalizes to prompt design: the single best signal in the model-convergence family wins on both prompts, though \emph{which} convergence signal is best depends on prompt structure. (ii)~OR-combining never beats the best singleton across the $381$ rules in either prompt regime. Dropping the confidence request entirely leaves the rule unaffected ($\pm 1$ F1), since TASR uses only the logit margin and answer text.

\begin{figure}[t]
  \centering
  \includegraphics[width=0.85\textwidth]{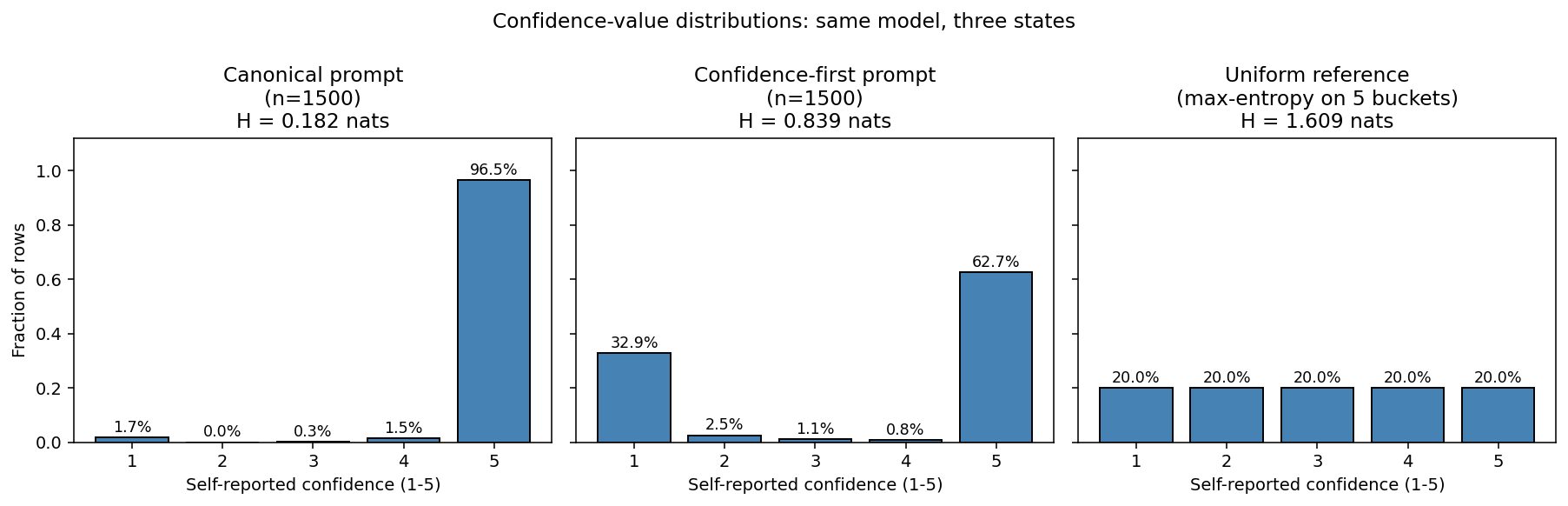}
  \caption{Verbalized confidence distributions under two prompt variants and the uniform reference. The canonical prompt collapses to $96.5\%$ mass on confidence $= 5$ (entropy $0.182$ nats); the confidence-first variant spreads to $62.7\%$ ($0.839$ nats) but remains far below the $1.609$-nat
  maximum.}
  \label{fig:collapse}
  \end{figure}

\subsection{Question-Type Breakdown}
\label{sec:qtype}

HotpotQA labels each question as \emph{bridge} (multi-hop chained facts) or \emph{comparison} (parallel facts about two entities). On the Qwen eval split ($241$ bridge, $59$ comparison), the answer-stability signal adapts to difficulty implicitly: on bridge questions it averages $2.59$ calls and gains $+2.05$ F1 over fixed-$k{=}3$; on comparison questions it stops earlier ($2.12$ calls) and ties on F1. The agent never receives the question-type label. This implicit adaptation is a consequence of the trajectory structure: comparison questions tend to stabilize earlier because the model can often resolve the comparison from the first two paragraphs, while bridge questions require chaining facts across paragraphs and take longer to converge.

\section{Discussion}
\label{sec:discussion}
\subsection{Why A One-Line Rule Works}
Answer stability is a \emph{trajectory} signal: it tests whether the model has stopped changing its mind, not whether retrieval has converged or whether confidence is high. The margin floor is a \emph{snapshot} guard: it withholds stopping when the model echoes a low-confidence answer. Every stopping decision reconstructs from two scalars per round, making it fully auditable without a separate probing setup.

\subsection{Why The Margin Is Necessary}
\label{sec:collapse}

Under the canonical prompt, $96.5\%$ of verbalized confidence values are $5$, yielding an entropy of $0.182$ nats (Figure~\ref{fig:collapse}). This is consistent across all three models, suggesting the collapse is a consequence of instruction-tuning and RLHF rather than a model-specific artifact. When isotonic calibration maps this degenerate distribution, it produces a single bucket: $\mathrm{conf}{=}5 \to \sim 0.57$ calibrated probability. The result is one effective category, useless for threshold rules.

The logit margin avoids this pathology because it is continuous (range $\sim 0$ to $14.9$ nats across all rounds on the Qwen tune split), so every training row contributes real variation to the PAV algorithm. Class-conditional separation is $2.75$ nats for the margin vs.\ $0.07$ nats for self-reported confidence, about a $40\times$ difference (Figure~\ref{fig:margin}). On this cell's Brier score, the margin achieves $0.199$ vs.\ $0.235$ for confidence and $0.246$ for the uninformative baseline: margin is $4\times$ better than confidence at beating the baseline ($0.047$ vs.\ $0.011$ Brier delta).

\subsection{Calibration Metrics Are Misleading}
\label{sec:brier}

\begin{sloppypar}
A recurring pattern in our ablations is that the calibrator with the lower Brier score does \emph{not} produce the better stopping rule (\S\ref{sec:calibration}): global isotonic wins on Brier ($0.192$ vs.\ $0.195$) but loses rule F1 by $1.1$, while Platt loses on Brier ($0.197$ vs.\ $0.195$) but wins rule F1 by $0.41$ on $12/18$ cells. The reason is structural: Brier measures average prediction accuracy across the whole input space, whereas rule F1 depends only on accuracy at the decision boundary, so a calibrator can have low average error yet blur the threshold region. We therefore recommend selecting calibrators by downstream rule metrics rather than by standard calibration scores.
\end{sloppypar}

\paragraph{Limitations.} The $100$-question tune split is small: each per-round isotonic calibrator is fit on only $100$ rows. The verbalized-confidence calibrator degenerates by round~$4$ (all values equal $5$) and falls back to that round's mean accuracy; TASR relies only on the margin calibrator, whose input retains full spread at every round. The evaluation covers short-answer factoid QA, both multi-hop and single-hop; whether the answer-stability signal generalizes to other RAG tasks (fact verification, long-form generation, conversational search) is untested. AS\_m25 also rests on, and is bounded by, several properties of our setup: access to top-$5$ token logprobs, a structured ``Answer:''-anchored response, temperature-$0$ decoding, and exact-match scoring of short answers. It may therefore not transfer to commercial APIs that hide logprobs, to long-form answers where a first-token margin carries little signal, or to tasks scored by semantic rather than exact match, which would also require refitting the isotonic map. The rule trades a fatter call-distribution tail (P95 of $4$ vs.\ $3$ for fixed-$k{=}3$) for a lower mean, which makes it unsuitable for latency-bound deployments where worst-case cost matters more than amortized cost. Devstral shows small regressions on open-domain ($-0.4$ to $-0.8$ F1), indicating that TASR's gains concentrate on weaker, overconfident models (Gemma); on strong single-pass models (Devstral) it is a near-tie that trims calls at no F1 cost. The Gemma overconfidence analysis (\S\ref{sec:gemma}) is observational; we show correlation between raw margin magnitude and accuracy loss, but do not identify a causal mechanism. We also do not benchmark TASR against a learned stopping controller; a small classifier trained on the same cached per-round signals is the natural comparison and is left to future work. Finally, TASR assumes a fixed ranked list revealed incrementally; extending it to agents that reformulate the query mid-trajectory remains future work.

\section{Conclusion}
\label{sec:conclusion}

TASR demonstrates that a single two-term predicate, which stops when the model repeats its previous answer and the calibrated logit margin exceeds one fixed threshold, suffices to beat fixed retrieval budgets at matched cost without any learned controller. On the primary three-model, two-dataset distractor grid it exceeds fixed-$k{=}3$ by $+3.42$ macro F1 at matched cost while retaining $94.8\%$ of fixed-$k{=}5$'s F1 at $62.6\%$ of its calls. The same rule and the same locked threshold transfer, with no per-cell tuning, to nine open-domain BM25 cells, nine dense-retrieval cells across two retriever families, and a $550$B production model roughly $18$--$23\times$ the design scale, with zero significant regressions throughout. That a calibrator fit once per model and then frozen survives all of these shifts indicates that the stopping signal is a property of the generation process itself rather than of the corpus, the retriever, or model size.

Two analyses explain why such a minimal rule suffices. Per-round isotonic regression maps each model's logit margins onto a common $P(\mathrm{EM})$ scale, which allows a single threshold to hold across architecturally distinct models and corrects the overconfident raw margins that otherwise make the weakest model appear most certain. This also explains why the gains concentrate where calibration matters most: the weak, overconfident model sees the largest improvements (Gemma: $+8.6$ distractor, $+1.7$ open-domain macro F1), whereas a well-calibrated model yields a matched-cost tie that reduces calls at no accuracy cost (Devstral). An exhaustive enumeration of $381$ candidate rules then shows that the rule's simplicity is discovered rather than assumed: OR-combining independently calibrated signals fires too eagerly, halting before evidence accumulates, whereas conjoining any further signal with the gate reduces to its single strongest constraint. The two retained terms are non-redundant by construction, since answer stability is a trajectory signal and the calibrated margin a per-round snapshot guard, so each guards against a failure mode the other misses; no richer combination Pareto-dominates them. Both results rest on a measurable pathology, the collapse of verbalized $1$--$5$ confidence on RLHF-tuned models, which the continuous logit margin avoids.

Since every stopping decision reconstructs from two scalars per round, TASR is fully auditable and serves as a training-free Pareto baseline against which learned stopping controllers can be measured. The most direct next step is that comparison, a small classifier trained on the same cached per-round signals, together with an assessment of whether the answer-stability signal extends beyond short-answer extractive QA; a controlled study that varies training data and RLHF reward while holding parameter count and serving precision fixed would further isolate the source of the overconfidence that calibration currently absorbs. We release the cached signals and evaluation harness so that the rule and competing alternatives can be reproduced and compared under identical conditions.

\subsubsection*{Disclosure of Interests.}
The authors have no competing interests to declare that are relevant to the content of this article.

\bibliographystyle{splncs04}
\bibliography{references}

\end{document}